# MODALITY EXCHANGE NETWORK FOR RETINOGENICULATE VISUAL PATHWAY SEGMENTATION

*Hua Han[1,2], Cheng Li[1], Lei Xie[3],*
*Yuanjing Feng[3], Alou Diakite[1,2] and Shanshan Wang[1,4]*

1. Shenzhen Institute of Advanced Technology, Chinese Academy of Sciences, Shenzhen, 518055, China
2. University of Chinese Academy of Science, Beijing, 100040, China
3. College of Information Engineering, Zhejiang University of Technology, Hangzhou 310023, China
4. Peng Cheng Laboratory, Shenzhen 518066, China

## ABSTRACT

Accurate segmentation of the retinogeniculate visual pathway (RGVP) aids in the diagnosis and treatment of visual disorders by identifying disruptions or abnormalities within the pathway. However, the complex anatomical structure and connectivity of RGVP make it challenging to achieve accurate segmentation. In this study, we propose a novel Modality Exchange Network (ME-Net) that effectively utilizes multi-modal magnetic resonance (MR) imaging information to enhance RGVP segmentation. Our ME-Net has two main contributions. Firstly, we introduce an effective multi-modal soft-exchange technique. Specifically, we design a channel and spatially mixed attention module to exchange modality information between T1-weighted and fractional anisotropy MR images. Secondly, we propose a cross-fusion module that further enhances the fusion of information between the two modalities. Experimental results demonstrate that our method outperforms existing state-of-the-art approaches in terms of RGVP segmentation performance.

## 1. INTRODUCTION

The segmentation of the retinogeniculate visual pathway (RGVP) is crucial for studying and understanding the anatomical development and progression of various diseases [1]. Additionally, visualizing RGVP through medical imaging techniques, such as magnetic resonance imaging (MRI), can provide valuable insights for surgical planning in order to address both intrinsic and extrinsic lesions along this pathway [2]. Accurate RGVP segmentation plays an important role in these applications. However, RGVP possesses a thin and elongated structure, making the accurate identification and differentiation of RGVP from neighboring tissues in MR images a challenging task.

To overcome these difficulties, researchers have delved into different RGVP segmentation techniques, and deep learning has emerged as a promising solution. By leveraging the power of deep learning algorithms, more precise and reliable RGVP segmentation results can be achieved, aiding in the accurate analysis and diagnosis of visual pathway disorders. For example, Mansoor et al. proposed an anterior visual pathway (AVP) segmentation method, which employs deep learning techniques to accurately identify the shape model of the optic nerve by integrating various MRI sequences including T1-weighted (T1w) imaging, T2-weighted (T2w) imaging, and fluid-attenuated inversion recovery (FLAIR) [3]. Tang et al. proposed an innovative deep learning method that integrates pixel-level and segmentation-level information using advanced integration techniques [4]. Ai et al. introduced a spatial probabilistic distribution map-based two-channel 3D U-net for visual pathway segmentation [5]. The majority of previous studies have focused on AVP segmentation using single-modal data. Although Mansoor et al. adopted three MRI sequences in their study, the three sequences are all structural imaging techniques. Incorporating extra functional imaging modalities can bring additional information to further improve the segmentation performance.

In view of this, Li et al. proposed the TPSN network [6], which integrates both structural T1w and functional fractional anisotropy (FA) images for the first time for AVP segmentation. In order to effectively exploit the different information provided by T1w MRI and diffusion MRI, Xie et al. developed a deep multi-modal fusion network for RGVP segmentation [7]. While promising performance has been reported, these existing studies still have limitations when it comes to multi-modal information extraction and fusion. Furthermore, they tend to employ conventional convolutional neural networks (CNNs) in their study that focus on extracting local features while neglecting the importance of global context information [8]. As a result, there is still plenty room for improvement in the field of RGVP segmentation.

In this study, we propose a modality exchange network (ME-Net) for segmentation of RGVP using multi-modal MR images. Specifically, ME-Net utilizes a channel and spatial hybrid attention module to achieve soft information swapping between T1w and FA images, so that only useful features can be kept and enhanced. Meanwhile, a cross-fusion module is designed to further enhance the complementary information fusion. Experiments were conducted on an open-source dataset, and the proposed ME-Net achieves better results than state-of-the art approaches.

## 2. METHODS

### 2.1. Overall Architecture

Figure 1 depicts the architecture of proposed ME-Net, comprising an exchange encoding module, a cross-fusion module, and a decoder [7]. Our key contributions lie in the exchange encoding and cross-fusion modules. Here, we will introduce the details of these modules.

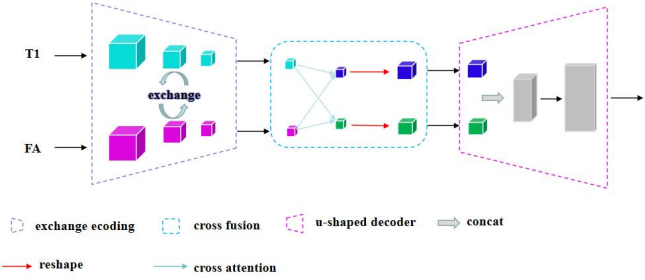

**Figure 1.** Illustration of the proposed ME-Net, which consists of an exchange encoding module, a cross-fusion module, and a decoder.

### 2.2. Exchange Encoding

The exchange encoding module aims to enhance the feature extraction process while minimizing the amount of introduced parameters. To achieve this, we employ two data exchange mechanisms: the Fixed Exchange Module (FEM) and the Adaptive Exchange Module (AEM) (Figure 2).

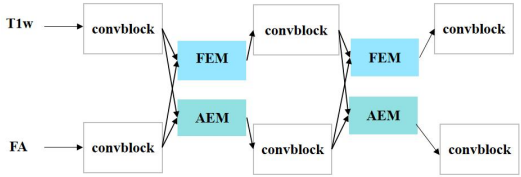

**Figure 2.** Details of the proposed exchange encoding module

#### 2.2.1 Fixed Exchange Module

In FEM, we employ a parameter-free attention mechanism, SimAm [10]. With SimAM, we obtain the coefficient maps in the channel dimension. With these coefficient maps, we determine the percentage of data to be retained and replace the remaining data using the data from the other modality accordingly. The detailed steps of FEM are illustrated in Figure 3.

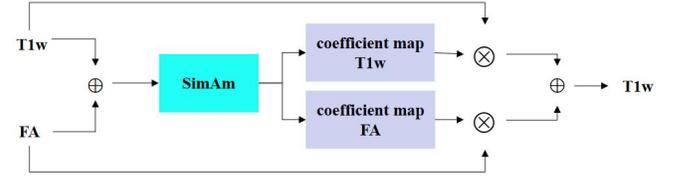

**Figure 3.** Details of FEM

In the work of Yang et al., the authors simulate the spatial suppression mechanism in human visual neurons to estimate the importance of individual neurons, thereby achieving the spatial channel attention [9]. In order to accurately evaluate the exchanged and retained features, we first overlay the data from T1w and FA:

$$X_{input} = X_{T1w} + X_{FA} \quad (1)$$

where $X_{T1} \in R^{C \times H \times W}$ represents the features extracted from the T1w modality, and $X_{FA} \in R^{C \times H \times W}$ represents the features extracted from the FA modality. Subsequently, we input the obtained feature maps into a SimAm attention block. Inspired by [11], here, we adopt a similar approach to generate the desired weight map, using $X_{input}$ as the input:

$$e_t = \frac{1}{M-1}\sum_{i=1}^{M-1}\left(-1 - (\omega_t x_{input_i} + b_t)\right)^2 + \left(1 - (\omega_t x_{input_t} + b_t)\right)^2 + \lambda \omega_t^2 \quad (2)$$

where $x_{input_t}$ and $x_{input_i}$ refer to the features of interest and other features, respectively, within a specific channel of the input feature map $X_{input}$. $M = H \times W$ denotes the total number of features in that channel. We can obtain the solution for $\omega_t$ and bias $b_t$ in Eq. (2) as:

$$\omega_t = -\frac{2(x_{input_t} - \mu_t)}{(x_{input_t} - \mu_t)^2 + 2\sigma^2 + 2\lambda} \quad (3)$$

$$b^t = -\frac{1}{2}(x_{input_t} + \mu_t)\omega_t \quad (4)$$

Here, $\mu_t = \frac{1}{M-1}\sum_{i=1}^{M-1} x_{input_i}$ and $\sigma_t^2 = \frac{1}{M-1}\sum_{i}^{M-1}(x_{input_i} - \mu_t)^2$ are the mean and variance of all features in the channel except for $x_{input_t}$. Given that the solutions presented in Eq. (3) and Eq. (4) are derived from a single channel, it is a valid assumption that all pixels within that channel adhere to a uniform distribution. Consequently, the minimal energy can be determined using the following equation:

$$e_t^* = \frac{4(\hat{\sigma}^2 + \lambda)}{(x_{input_t} - \hat{\mu})^2 + 2\hat{\sigma}^2 + 2} \quad (5)$$

where $\hat{\mu} = \frac{1}{M}\sum_{i=1}^{M} x_{input_i}$, and $\hat{\sigma}^2 = \frac{1}{M}\sum_{i=1}^{M}(x_{input_i} - \hat{\mu})^2$. In Eq. (5), a lower energy value $e_t^*$ indicates that the feature $x_{input_t}$ is more distinct from the surrounding features. Hence, the importance of each feature can be determined as $1/e_t^*$. We use E groups all $e_t^*$ across channel and spatial dimensions. Afterwards, we use $1/E$ to compute the coefficient map of T1w:

$$F_{T1w} = sigmoid(\frac{1}{E}) \quad (6)$$

Similarly, we can obtain the coefficient map of FA:
$$F_{FA} = 1 - F_{T1w} \quad (7)$$
And the output of FEM for T1w modality becomes:
$$X_{T1} = F_{T1w}X_{T1w} + F_{FA}X_{FA} \quad (8)$$

### 2.2.2 Adaptive Exchange Module

For FA, we designed another lightweight and learnable exchange module, AEM, where we combined efficient channel attention (ECA) with spatial attention mechanisms (Figure 4).

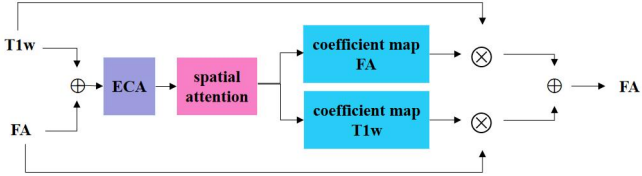

**Figure 4.** Details of AEM

In AEM, the ECA module effectively captures the dependencies and interactions between different channels within the network. This allows for more accurate and informative feature representations.

Following what we have done for T1w modality, we overlay the T1w modality and FA modality to generate the input to ECA :
$$X_c = \sigma(f(GAP(X_{input}))) \quad (9)$$
where $X_{input} \in R^{C \times H \times W}$ is the input feature map. GAP stands for global average pooling. f (X) represents a 1D convolutional layer. σ represents the sigmoid function. Then, we apply a channel-wise attention to it:
$$X_F = concat(AP(X_c), MP(X_c)) \quad (10)$$
$$F_{FA} = sigmoid(f(X_F)) \quad (11)$$
where AP and MP represent average pooling and max pooling, respectively. f(x) represents a 2D convolution layer. We concatenate them together and use another 2D convolution layer for channel compression. Finally, we apply a sigmoid function to generate a coefficient map $F_{FA}$. Similarly, we can obtain the coefficient map for T1w as:
$$F_{T1w} = 1 - F_{FA} \quad (12)$$
Finally, we obtain the output result of the FA modality:
$$X_{FA} = F_{FA}X_{FA} + F_{T1w}X_{T1w} \quad (13)$$

### 2.3. Cross-Fusion Module

To further enhance the multi-modal information fusion, we incorporate a cross-fusion module in our framework that leverages a powerful cross-attention mechanism to match and fuse information from different modalities, allowing for comprehensive integration and capturing of both local and global contextual dependencies.

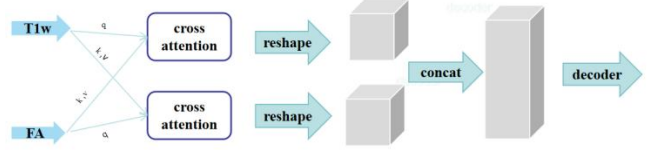

**Figure 5.** The proposed cross-fusion module

The cross-fusion module generates cross attention maps for each pair of class features and query sample features so as to highlight the target object regions, making the extracted features more discriminative [12]. In CrossViT, the cross attention module is utilized for multi-scale feature fusion [13]. The operations of the cross-fusion module can be expressed as:
$$\text{CrossAttention}_{T1w} = \text{softmax}\left(\frac{Q_{FA}K_{FA}^T}{\sqrt{d_k}}\right)V_{T1} \quad (14)$$
$$\text{CrossAttention}_{FA} = \text{softmax}\left(\frac{Q_{T1}K_{T1}^T}{\sqrt{d_k}}\right)V_{FA} \quad (15)$$
$$y = concat(reshape(\text{CrossAttention}_{T1w}),$$
$$reshape(\text{CrossAttention}_{FA})) \quad (16)$$

In our approach, we transform the FA modality through three linear layers to obtain the $K_{FA}$, $Q_{FA}$, and $V_{FA}$ representations. Simultaneously, we apply the same linear transformations to the T1w modality, resulting in the corresponding $K_{T1}$, $Q_{T1}$, and $V_{T1w}$ representations.

Then, we input the $K_{FA}$ and $Q_{FA}$ representations from the FA modality, along with the $V_{T1w}$ representation from the T1w modality, to a cross-attention mechanism. After that, we obtain the $\text{CrossAttention}_{T1w}$ representation. We follow the same procedure to obtain the $\text{CrossAttention}_{FA}$ representation.

### 2.4. Decoder and Segmentation Output Generation

The decoder is a classical decoder of the U-shaped network. It receives the features from the cross-fusion module and generates the RGVP segmentation results for consideration.

## 3. EXPERIMENTS

### 3.1. Dataset

We used an open-source dataset from the human connectome project (HCP), which consists of data from 102 cases. The dataset provides both T1w structural MRI and preprocessed diffusion MRI (dMRI) data. The matrix size of the data is 145×174×145, with a voxel resolution of 1.25×1.25×1.25 mm³. We generated the reference data for the RGVP segmentation in 102 cases by mapping the streamlines onto voxel-based binary images. In all experiments, we divided the dataset of 102 cases into training, validation, and testing sets using a ratio of 8:1:1.

### 3.2. Implementation Details and Evaluation Metrics

We used an NVIDIA GeForce RTX 2080 GPU to run the experiments. Each model was trained for 200 epochs with a batch size of 40. The initial learning rate during training was 0.00015, with a weight decay of 0.00001. We utilized a combination of Dice loss and cross-entropy loss for network training:

$$\text{BCELoss}(y, t) = y * \log(y') + (1 - y) * \log(1 - y') \quad (17)$$

$$\text{DiceLoss}(y, t) = 1 - \frac{2*|y' \cap y|}{|y'| + |y|} \quad (18)$$

$$\text{Loss}(y', y) = \text{BCELoss}(y', y) + \text{DiceLoss}(y', y) \quad (19)$$

where y' represents the network prediction and y represents the label. We compared the experimental results with Unet [14], Unet++ [15], TPSN [6], and Deep Multimodal Fusion Network [7]. Dice similarity coefficient (DSC), relative absolutevolume difference (RAVD), Hausdorff distance (HD), and average symmetric surface distance (ASSD) are calculated to evaluate the segmentation performance of different methods.

$$\text{DSC} = \frac{y \cap y'}{y + y'} = \frac{2TP}{2TP + FP + FN} \quad (20)$$

where TP denotes true positive predictions. FP denotes false positive predictions. FN denotes false negative predictions.

$$\text{RAVD} = \left| \frac{V_{seg}}{V_{gt}} - 1 \right| \times 100\% \quad (21)$$

where $V_{seg}$ represents the segmentation volume, and $V_{gt}$ represents the label volume.

$$\text{HD} = \max(h(A, B), h(B, A)) \quad (22)$$

In this context, A and B represent the sets of groundtruth segmentation and predicted segmentation, respectively. h(A, B) denotes the one-way Hausdorff distance, which is defined as the maximum of the minimum distances between each point from A to B. Similarly, h(B,A) is called the one-way Hausdorff distance from set B to set A.

$$\text{ASSD} = \frac{1}{|S_a| + |S_b|} \left( \sum_{a \in S_a} d(a, S_b) + \sum_{b \in S_b} d(b, S_a) \right) \quad (23)$$

Where $S_a$ and $S_b$ represent the model prediction results and ground truth, respectively.

### 3.3. Results

#### 3.3.1 Quantitative results

We conducted a thorough validation of our proposed ME-Net by comparing its results with state-of-the-art (SOAT) methods. Quantitative results are listed in Table 1. Our network exhibited significant improvements over the SOTA methods. Specifically, the average dice value was nearly one point higher, RVD was 4.53% lower, HD was 0.4 lower, and ASD was 0.008 lower than the current best method. Additionally, our proposed method achieved a dice score that was 0.8% better than the current best method, demonstrating the effectiveness of our network architecture.

Table 1: Quantitative results of different methods. All results of the comparison method are adopted from the original papers.

| Methods | DSC (%) | RAVD (%) | HD (mm) | ASSD (mm) |
|---|---|---|---|---|
| Unet | 83.3 | -- | 2.913 | 0.197 |
| Unet++ | 84.3 | -- | 3.066 | 0.168 |
| TPSN | 85.5 | -- | 2.330 | 0.162 |
| (Xie et al. 2023) | 87.4 | 7.762 | 2.679 | 0.135 |
| Ours | 88.2 | 3.232 | 2.278 | 0.127 |

#### 3.3.2 Visualization results

Figure 6 plots the segmentation maps of our method and one existing SOTA method. We can observe that qualitatively, our method can still achieve better segmentation results than the comparison method. Our method can better preserve the fine details of the target, while the comparison segmentation method shows some detail loss. These visualization results further validate the effectiveness of our proposed ME-Net for our task of RGVP segmentation.

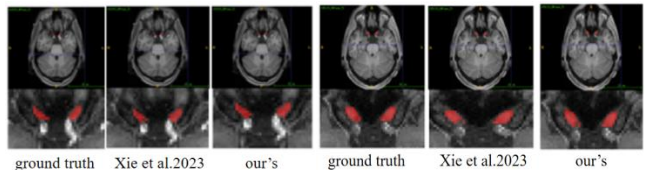

**Figure 6.** Example visualization results

### 4. CONCLUSION

In this work, we propose a modality exchange network, ME-Net, for RGVP segmentation. ME-Net has two key contributions: multi-modal information exchange encoding and cross fusion. The exchange encoding module uses a hybrid attention mechanism and soft swapping method to enhance feature extraction by exchanging information across modalities. The cross-fusion module employs a cross attention mechanism to better capture global information during multi-modal fusion. Experiments on the open-source HCP dataset validated the effectiveness of ME-Net for RGVP segmentation. This advancement in RGVP segmentation can play a vital role in the study and comprehension of the anatomical development and progression of various visual-related diseases.

### 5. COMPLIANCE WITH ETHICAL STANDARDS

## 6. ACKNOWLEDGMENTS

This research was partly supported by the National Natural Science Foundation of China (62222118, U22A2040), Guangdong Provincial Key Laboratory of Artificial Intelligence in Medical Image Analysis and Application (2022B1212010011), Shenzhen Science and Technology Program (RCYX20210706092104034, JCYJ20220531100213029), and Key Laboratory for Magnetic Resonance and Multimodality Imaging of Guangdong Province (2023B1212060052).

# References

[1] R.L. Harrigan, Swetasudha Panda, A.J. Asman, K.M. Nelson, Shikha Chaganti, M.P. DeLisi, B.C.W. Yvernault, S.A. Smith, R.L. Galloway L.A. Mawn and B.A. Landman, "Robust optic nerve segmentation on clinically acquired computed tomography," Journal of Medical Imaging, 1(3), 034006-034006.

[2] Jianzhong He, Fan Zhang, Guoqiang Xie, Shun Yao, Yuan jing Feng, D.C.A. Bastos, Yogesh Rathi, Nikos Makris, Ron Kikinis,A.J. Golby and L.J. O'Donnell, "Comparison of multiple tractography methods for reconstruction of the retinogeniculate visual pathway using diffusion MRI," Human Brain Mapping, 42(12), 3887-3904.

[3] Awais Mansoor, J.J. Cerrolaza, Rabia Idrees, Elijah Biggs, M.A. Alsharid, R.A. Avery, M.G. Linguraru, "Deep learning guided partitioned shape model for anterior visual pathway segmentation," IEEE transactions on medical imaging, 35(8), 1856-1865.

[4] Xue-song Tang, Hui Wei, Kuangrong Hao, Mingbo Zhaoa, and Dawei Li, "Integrating pixels and segments: A deep-learning method inspired by the informational diversity of the visual pathways," Neuro computing, 396, 314-323.

[5] Danni Ai, Zhiqi Zhao, Jingfan Fan, Hong Song, Xiaoxia Qu, Junfang Xian and Jian Yang, "Spatial probabilistic distribution mapbased two-channel 3D U-net for visual pathway segmentation," Pattern Recognition Letters, 138, 601-607.

[6] Siqi Li, Zan Chen, Wenlong Guo, Qingrun Zeng, and Yuanjing Feng, "Two parallel stages deep learning network for anterior visual pathway segmentation," In Computational Diffusion MRI: International MICCAI Workshop, Lima, Peru, October 2020 (pp. 279-290). Cham: Springer International Publishing.

[7] Lei Xie, Lin Yang, Qingrun Zeng, Jianzhong He, Jiahao Huang, Yuanjing Feng,Evgeniya Amelina and Mihail Amelin**,** "Deep Multimodal Fusion Network for the Retinogeniculate Visual Pathway Segmentation," In The 42nd Chinese Control Conference.

[8] Yikang Ding, Wentao Yuan, Qingtian Zhu, Haotian Zhang, Xiangyue Liu, Yuanjiang Wang and Xiao Liu, "Transmvsnet: Global context-aware multi-view stereo network with transformers," In Proceedings of the IEEE/CVF Conference on Computer Vision and Pattern Recognition (pp. 8585-8594).

[9] Lingxiao Yang, Ru-Yuan Zhang, Lida Li and Xiaohua Xie, "Simam: A simple, parameter-free attention module for convolutional neural networks," In International conference on machine learning (pp. 11863-11874). PMLR

[10] Qilong Wang, Banggu Wu, Pengfei Zhu, Peihua Li, Wangmeng Zuo and Qinghua Hu, "ECA-Net: Efficient channel attention for deep convolutional neural networks," In Proceedings of the IEEE/CVF conference on computer vision and pattern recognition (pp. 11534-11542).

[11] B.S. Webb, N.T. Dhruv, S.G. Solomon, Chris Tailby and Peter Lennie1, "Early and late mechanisms of surround suppression in striate cortex of macaque," Journal of Neuroscience, 25(50), 11666-11675.

[12] Ruibing Hou, Hong Chang, Bingpeng Ma, Shiguang Shan and Xilin Chen, "Cross attention network for few-shot classification," Advances in neural information processing systems, 32.

[13] Chun-Fu(Richard) Chen, Quanfu Fan and Rameswar Panda, "Crossvit: Cross-attention multi-scale vision transformer for image classification," In Proceedings of the IEEE/CVF international conference on computer vision (pp. 357-366).

[14] Olaf Ronneberger, Philipp Fischer, and Thomas Brox, "U-net: Convolutional networks for biomedical image segmentation," In Medical Image Computing and Computer-Assisted Intervention–MICCAI 2015:18th International Conference, Munich, Germany, October 5-9, 2015, Proceedings, Part III 18 (pp. 234-241). Springer International Publishing.

[15] Zongwei Zhou, Md Mahfuzur Rahman Siddiquee, Nima Tajbakhsh and Jianming Lian,"Unet++: Redesigning skip connections to exploit multiscale features in image segmentation," IEEE transactions on medical imaging, 39(6), 1856-1867.